\documentclass{ws-ijmpcs}
\usepackage{hyperref}
\hypersetup{colorlinks=true,urlcolor=black,linkcolor=red,citecolor=blue}

\begin{document}

\markboth{V. Gaibler}
{Asymmetries of AGN jets in inhomogeneous media}

%
\catchline{}{}{}{}{}
%

\title{ASYMMETRIES OF AGN JETS IN INHOMOGENEOUS MEDIA
}

\author{VOLKER GAIBLER}

\address{Universit\"at Heidelberg, Zentrum f\"ur Astronomie, Institut f\"ur Theoretische Astrophysik,\\
Albert-Ueberle-Str. 2, 69120 Heidelberg, Germany\\
v.gaibler@uni-heidelberg.de}

\maketitle


\begin{abstract}
  Considerable asymmetries in jets from active galactic nuclei (AGN) and associated double radio sources can be caused by an inhomogeneous interstellar medium of the host galaxy. These asymmetries can easily be
  estimated by 1D propagation models, but hydrodynamical simulations have shown that the actual asymmetries can be considerably larger. With a set of smaller-scale hydrodynamical simulations we examine these asymmetries, and 
  find they are typically a factor of $\sim 3$ larger than in 1D models. We conclude that, at high redshift, large asymmetries in radio sources are expected in gas-rich galaxies with a clumpy interstellar medium.
\end{abstract}

\ccode{PACS numbers: 98.54.Gr, 98.58.Ay, 98.62.Nx}

\section{Introduction}

The propagation of jets from active galactic nuclei (AGN) is strongly dependent on the medium they are moving through. Outside the host galaxy, at scales of more than some 10 kpc, this is a mostly diffuse medium with rather small density gradients. At smaller scales, however, the jets interact with the interstellar medium (ISM) of the host galaxy, which is a multi-phase medium with a large range of densities and temperatures, and it has a very clumpy and filamentary structure in particular at high densities (cold neutral atomic and molecular phase). Our hydrodynamic simulations (Ref.~\refcite{Gaibler+2011}, hereafter referred to as Paper I) have shown that jets and the jet cocoon can exhibit pronounced asymmetries in presence of a massive gaseous disk as is expected for galaxies at high redshifts. Any clouds in the jets path cause a strong thermalization and lead to the formation of a continuously driven blast wave for each of the two jets in the central region of the disk. These blast waves will merge soon 
and expand 
roughly spherically until they 
vertically break out of the disk. In contrast to the blast waves, however, the propagation of the two jet beams depends on the inertia of obstructing ambient matter and any asymmetry in the stochastically located clumps results in an asymmetry of the radio source\cite{Pedelty+1989,McCarthy+1991,ArshakianLongair2000,Thomasson+2003,Jeyakumar+2005,Orienti+2007}. Once both jets have broken out of the ISM, they will propagate more efficiently in the diffuse ambient gas but still keep an imprint of this initial asymmetry, which can also be regarded as a propagation delay between both jets for the large-scale propagation. We have derived simple method that can be applied to any (statistically defined) ISM structure and results in a probability distribution function (PDF) for the asymmetries and hence for the associated radio emission. Assuming jet propagation according to one-dimensional momentum balance, this method was able to describe the asymmetries in the hydrodynamical simulation to first order, but it was 
found that 
considering the full 
hydrodynamics the actual delay is even larger due to ``second order'' effects: The blast wave already affects the ISM structure, moves and compresses clumps, and hence changes the density profile that the jet needs to propagate through. For the specific hydrodynamical setup of Paper I, the expected delay time between the jets from the momentum balance Monte Carlo runs was 0.7 Myr, but the actually measured delay was 2.7 Myr. 

Since the 3D hydrodynamical simulations mentioned are computationally expensive, only one simulation run was performed and analyzed in Paper I and the question had to remain open whether this difference by a factor of more than 3 is typical or just a stochastic realization at the upper end of the delay distribution (with a probability of less than 1 percent). By imposing restrictive adaptive mesh refinement (AMR) criteria on the simulations, we were able to run eight additional simulations for the early stage of propagation and will present those in this contribution. This allows a better judgment of the asymmetries found in hydrodynamical simulations compared to the 1D estimate. Due to the stochastic nature of the clump-related asymmetries, only probabilities can be derived rather than specific values -- within this model, it is not possible to determine whether clumps or filaments were present at some location in the past. Furthermore, it is important to keep in mind that the asymmetries discussed here are 
only due to asymmetries in the ISM, while in actual radio sources observed asymmetries may also be caused by ambient gas asymmetries on larger scales\cite{Jeyakumar+2005} or other effects\cite{GopalKrishnaWiita2004}, including light travel times\cite{Scheuer1995}, Doppler boosting\cite{WardleAaron1997} and intrinsic instabilities of the jet beam\cite{ONeill+2012}.

\section{Simulation setup}

The simulations described in the following were performed with the setup as in Paper I, with the differences limited to the ones described here. A powerful jet of kinetic power $5.5 \times 10^{45}$ erg s$^{-1}$ was injected at $t = 2$ Myr in the center of a clumpy gaseous disk of mass $10^{11} \, M_\odot$. To reduce the computational costs and hence allow for multiple simulations, strict AMR criteria were imposed: in the previous simulation, refinement was triggered within the disk and by density and pressure gradients between cells larger than 10 percent, which resulted in a refinement to the maximum resolution of 62.5 pc for virtually the entire affected system. In the new setup, this maximum refinement was limited to $< 5$ kpc distance from the jet axis and 6 kpc vertically from the disk midplane, outside of which the resolution gradually drops to 1 kpc. Also for lower resolution regions, momentum conservation is ensured by the code, but propagation will deviate there since density averages 
are used for these cells. The propagation through the disk, however, is fully resolved and hence has the same precision as the fully resolved run in Paper I. Eight of these small-scale simulations are performed, each with a different realization of clumpy disk generated in Fourier space. Accordingly, the global disk properties as the density PDF and disk mass are unchanged, but the spatial distribution of the clumps varies stochastically. Although eight runs statistically are still a very limited set, they allows us to test the conjecture that the larger asymmetries in hydrodynamical runs compared to 1D estimates are typical rather than an exception.

\section{Results}

\begin{figure}[tb]
\centerline{\psfig{file=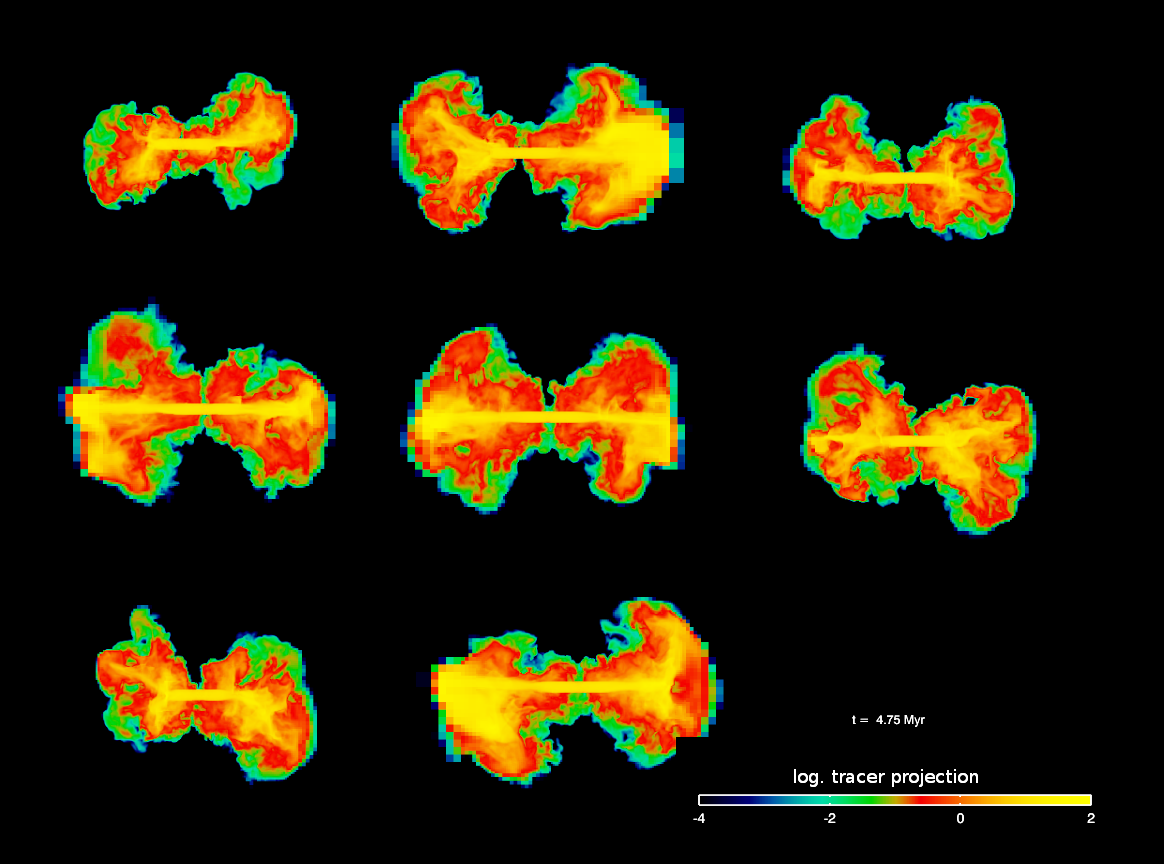,width=\textwidth}}
\vspace*{8pt}
\caption{Projection of the jet plasma tracer field for the eight runs at a
common time $t = 4.75$ Myr. This shows the interaction of the jet beam with the
clumpy ISM and gives an impression of the morphology of the radio source, since
only the jet-originated matter will emit synchrotron radiation. The projection
is displayed with a logarithmic scale.
\label{f1}}
\end{figure}
The differences of the eight different realizations of the clumpy disk structure are immediately evident from Fig.~\ref{f1}. The total extent of the cocoon and the blast wave is quite similar. The morphology and the location of the jet head, however, are very dependent on the actual clump locations. In some cases the jet propagates more easily and already forms prominent lobes (e.g. right jet of \#2 or left jet of \#8), in other cases it obviously hits a massive cloud and temporarily gets deflected, the cocoon then expanding sideways (e.g. left jet of \#1 or right jet of \#8).

\begin{figure}[tb]
\centerline{\psfig{file=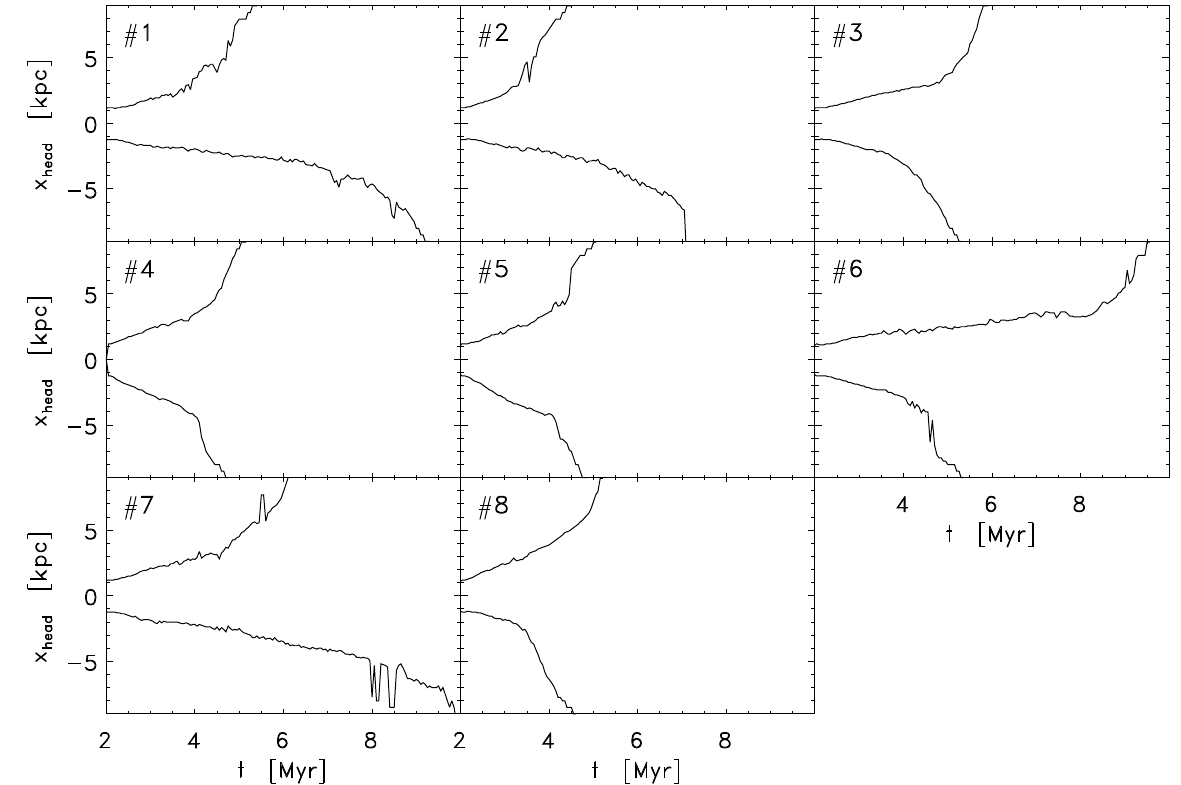,width=\textwidth}}
\vspace*{8pt}
\caption{Propagation of the jet head on the jet axis for the eight runs, with
the two jets of one simulation shown for each plot in the panel at negative and
positive coordinates $x_\mathrm{head}$. The delay for each run is determined
from the times when the jet heads reach a certain distance, outside the vertical
extent of the initial disk. The delays discussed in the text use a fiducial location $x_\mathrm{head} = \pm 5$ kpc.
\label{f2}}
\end{figure}
To measure the propagation delay between both jets, we have plotted the position of the jet head over time in Fig.~\ref{f2} for all simulations. The location of the jet head was determined from the axial extent of the jet tracer field, based on the projection of a 1 kpc wide region around the jet axis. Spikes as for simulation \#7 at $t = 8$ Myr result from the cocoon expanding around a clump and reaching towards the axis behind it, and are only a consequence of the definition of the jet head location. Shortly after, the jet beam moves forward since the cloud was pushed away and matter ablated off its surface. We measured the delay times between both jets by taking the times when the jet head reaches a position of $\pm 5$ kpc. 
The mean delay is $1.9$ Myr, the percentiles used in Paper I are $\Delta t_{50\%} = 1.8$ Myr, $\Delta t_{75\%} = 3.4$ Myr and $\Delta t_{90\%} = 4.3$ Myr. Although these numbers are strongly affected by the small sample size, they are all consistently larger than the 1D estimate by a factor of $\approx 3$. The numbers are only weakly dependent on the distance at which this delay is measured. 

\section{Conclusions}

Our hydrodynamical simulations of the jet propagation through the clumpy disk suggest that the real asymmetries between both jets are typically a factor of $3$ larger than the estimates derived from the Monte-Carlo 1D method described in Paper I. This is a result of the complex interplay between the jet and its environment -- with the formed blast wave partially shaping the medium the jet is propagating through. Our results show that the simulation of Paper I shows a rather typical asymmetry for the chosen parameters, not an exceptionally large one. Furthermore, our suggestion that observed gas masses in galaxies may considerably contribute to asymmetries in observed 3CRR\cite{Laing+1983} radio galaxies is further strengthened by these findings. While the stochastic nature of these asymmetries does not allow to derive firm numbers for individual sources, it allows for the determination of a typical range of asymmetries that should be expected. For the thick gaseous disks and large gas masses conjectured to 
be present in massive galaxies a redshifts $\gtrsim 2$, strong asymmetries are expected in the radio sources. Also, the considerable difference between the hydrodynamical and 1D results show that the propagation of jets if far more than only piercing through ISM of the host galaxy. At early times, there is plenty of interaction with the ISM, mostly via the blast wave that is formed already at very small scales (cf. the scale $L_{1b}$ in Ref.~\refcite{Krause+2012}). And although we clearly see the effects of the clumpy ISM in our simulations, the actual dense structures in the ISM as molecular clouds and filaments are still below our resolution limit. The smaller-scale simulations of Ref.~\refcite{Wagner+2012}, however, show that the interaction is qualitatively similar and insensitive to the cloud volume filling factor, but dependent on the maximum size of the clouds. The interaction with the ISM and feedback on the star formation of the galaxy here clearly depends on the actual ISM properties. Also at later times, the over-pressured cocoon and its turbulent motions pressurize and change the ISM of the host galaxy, providing feedback on the star formation\cite{Gaibler+2012} -- but at these times, the jets have already moved out far beyond the extents of ISM of the host galaxy. The persisting asymmetries in the radio source, however, may still reflect the strong interaction during its early evolution.

\section*{Acknowledgments}

VG wishes to acknowledge financial support by the German Research Foundation
(DFG) through Priority Programme SPP 1177 and in part by Sonderforschungsbereich
SFB 881 ``The Milky Way System'' (subproject B4). Computations were performed on
the SFC cluster of TMoX at MPE and the Power6 VIP at RZG Garching.


\end{document}